\documentclass[%
 aip,
rsi,%
 amsmath,amssymb,
 reprint,%
]{revtex4-1}

\usepackage[dvipsnames]{xcolor}
\usepackage{graphicx}
\usepackage{dcolumn}
\usepackage{bm}
\usepackage{lipsum}

\usepackage{dcolumn}
\usepackage{epstopdf}
\usepackage{xcolor}
\usepackage{natbib}
\usepackage{subcaption} 
\usepackage{float}
\usepackage{tabularx}
\usepackage{mathrsfs}
\usepackage{upgreek}
\usepackage[a]{esvect}
\setcitestyle{square}
\usepackage[toc,page]{appendix}
\usepackage[font={small,normalfont}]{caption}
\usepackage[colorlinks=true, allcolors=blue]{hyperref}
\usepackage[normalem]{ulem}
\usepackage[capitalise]{cleveref}

\newcommand{\micron}{{\upmu \mbox{m}}}

\begin{document}


\title{Efficient backward x-ray emission in a finite-length plasma irradiated by a laser pulse of ps duration}

\author{I-L. Yeh}
\affiliation{Department of Physics, University of California San Diego, La Jolla, CA 92093}

\author{K. Tangtartharakul}
\affiliation{Department of Mechanical and Aerospace Engineering, University of California San Diego, La Jolla, CA 92093}

\author{H. Tang}
\affiliation{Center for Ultrafast Optical Science, University of Michigan, Ann Arbor, Michigan 48109}

\author{L. Willingale}
\affiliation{Center for Ultrafast Optical Science, University of Michigan, Ann Arbor, Michigan 48109}

\author{A. Arefiev}
\email[]{aarefiev@ucsd.edu}

\affiliation{Department of Mechanical and Aerospace Engineering, University of California San Diego, La Jolla, CA 92093}

\affiliation{Center for Energy Research, University of California San Diego, La Jolla, CA 92093}

\date{\today}

\begin{abstract}
Motivated by experiments employing ps-long, kilojoule laser pulses, we examined x-ray emission in a finite-length underdense plasma irradiated by such a pulse using two dimensional particle-in-cell simulations. We found that, in addition to the expected forward emission, the plasma also efficiently emits in the backward direction. Our simulations reveal that the  backward emission occurs when the laser exits the plasma. The longitudinal plasma electric field generated by the laser at the density down-ramp turns around some of the laser-accelerated electrons and re-accelerates them in the backward direction. As the electrons collide with the laser, they emit hard x-rays. The energy conversion efficiency is comparable to that for the forward emission, but the effective source size is smaller. We show that the ps laser duration is required for achieving a spatial overlap between the laser and the backward energetic electrons. At peak laser intensity of $1.4\times 10^{20}~\rm{W/cm^2}$, backward emitted photons (energies above 100~keV and $10^{\circ}$ divergence angle) account for $2 \times 10^{-5}$ of the incident laser energy. This conversion efficiency is three times higher than that for similarly selected forward emitted photons. The source size of the backward photons ($5~\micron$) is three times smaller than the source size of the forward photons. 
\end{abstract}

\maketitle


\section{Introduction} \label{sec: intro}

The proliferation of ultrahigh-intensity, short-pulse laser systems based on chirped-pulse amplification \cite{Strickland1985} has stimulated considerable interest in a wide range of laser-based applications. These include compact broadband all-optical sources of hard x-rays and gamma-rays that employ laser-irradiated plasmas~\cite{Ridgers2012PRL, Corde_ph,nakamura2012high, taphuoc.natphot.2012,Albert2016PPCF, Huang2016PRE, Gong2018PPCF, huang2019highly,feng.scirep.2019, wang_power_PRA2020, Shen2021APL,Guenther2022Naturecom,armstrong.frontier_in_phys.2023}. Hereafter, we refer to them simply as laser-driven x-ray sources. These sources can be used for various diagnostics like x-ray absorption spectroscopy~\cite{yu.apl.2018,kettle.prl.2019} and x-ray phase imaging~\cite{cole.scirep.2015,Albert2016PPCF,siegfried.revmodphys.2009,antonelli.2019.phaseimaging,hussein.scirep.2019,ostermayr.natcom.2020}. 

There is a great interest in using laser-driven x-ray sources to diagnose high-energy-density experiments~\cite{siegfried.revmodphys.2009,antonelli.2019.phaseimaging,hussein.scirep.2019}. These experiments are often performed at large laser facilities, which means that the laser beam needed for the x-ray source might be available by default or with minor modifications to the setup. This is an important advantage, but using an existing laser system for the development of a laser-driven x-ray source introduces a constraint. One must work with the available laser parameters, such as laser pulse duration and laser peak intensity, that define the key physics that we further discuss in \cref{sec-background}. In this context, PW-class, kJ-level, ps-long lasers represent an important category because they are available at multiple user facilities worldwide. Examples include OMEGA EP~\cite{omega-ep} at LLE (USA), ARC~\cite{nif-arc} at NIF (USA), PETAL~\cite{petal} at LMJ (France), and SG-II UP~\cite{nlhplp} at NLHPLP (China). 

This paper focuses on an unexpected feature of hard x-ray emission from an underdense plasma driven by a ps duration laser pulse with a peak intensity $\gtrsim 10^{20}$ $\rm{W/cm^2}$. Using fully kinetic two-dimensional (2D) particle-in-cell (PIC) simulations, we found that a finite-length plasma efficiently emits backward directed x-rays in addition to conventional forward-directed x-rays. We anticipate our results to be particularly informative to those researchers who are working on developing laser-driven x-ray sources using OMEGA EP~\cite{omega-ep} or SG-II UP~\cite{nlhplp} lasers.

We found that the backward emissions is associated with the laser exiting the plasma along a density down-ramp -- an aspect often omitted from computational research. As the laser travels along the density down-ramp, it creates large-scale charge separation. This is because a high-intensity laser tends to push plasma electrons forward. The charge-separation induces and maintains a longitudinal electric field in the forward direction. The field is sufficiently strong to turn around moderate-energy (compared to the peak energy) electrons and re-accelerate them backward. This movement in the opposite direction to the laser pulse propagation causes the electrons to efficiently emit backward-directed x-rays. The backward acceleration takes time, so the laser pulse duration has to be sufficiently long for the described scenario to play out. Otherwise, the backward-accelerated electrons reach their peak energy after the laser pulse is already gone.
For solid target interactions, this is a very well-known process and the refluxing electrons can significantly alter global processes, such as the target normal sheath acceleration (TNSA) of ion beams~\cite{Hatchett_PoP_2000}.
Previous experiments have observed ion acceleration from the underdense target sheath field in direct laser acceleration experiments~\cite{Willingale_PRL_2006}, illustrating that the field strength can be significant even in the longer density gradients associated with a gas jet target.

The remainder of the paper is organized as follows. \Cref{sec-background} provides additional background and introduces the field configuration that arises in the plasma irradiated by a ps-long laser pulse. \Cref{sec-estimation} provides estimates for synchrotron emission by electrons that we use in subsequent sections to understand our results. \Cref{sec: back_emission} presents simulation results for the $\gamma$-ray emission at the density down-ramp. \Cref{sec: backward electrons} explains the mechanism responsible for the generation of backward-moving electrons that emit backward-directed $\gamma$-rays. \Cref{sec: pulse duration} examines the impact of the laser pulse duration on the backward emission. \Cref{sec: source size} provides a variety of metrics  for the forward- and backward-directed $\gamma$-ray beams. \Cref{sec: summary} provides a brief summary of our results and a discussion.


\section{Background} \label{sec-background}

There are multiple concepts for laser-driven x-ray sources, but they all generally involve two key steps. The first step is the generation of ultra-relativistic electrons that occurs during the laser-plasma interaction. The second step is the emission of energetic photons by the generated electrons when the electrons experience strong acceleration by electric and/or magnetic fields. These fields can be either macroscopic, like the fields of the laser or the collective fields of the plasma, or microscopic, like the fields of individual ions or atoms. The two steps can happen sequentially or simultaneously. The mechanisms used for the implementation determine the conversion efficiency of the incident laser energy into the photons and the photon beam characteristics such as spectrum, photon number, and brilliance. 

The two most popular mechanisms for generating ultra-relativistic electrons are the laser-wakefield acceleration (LWFA)~\cite{tajima1979PRL,mangles.nature.2004,geddes.nature.2004,esarey.revmodphys.2009} and the direct laser acceleration (DLA)~\cite{pukhov1999DLA, Robinson_2013_PRL, Khudik-POP_2016, arefiev2016beyond, gong2020PRE, Jirka2020njp, hussein2021njp,cohen.sciadv.2024,babjak.prl.2024,tang.njp.2024}. They have opposing requirements on laser pulse duration $\tau_l$ with respect to the electron response time $\tau_e$. In LWFA, the electrons are accelerated by a longitudinal plasma electric field that moves forward just behind the laser pulse. The laser pulse creates the moving plasma field structure if $\tau_l < \tau_e$. In DLA, the electrons gain their energy directly from the laser electric field~\cite{Khudik-POP_2016}. The DLA is particularly effective in the presence of quasi-static radial electric and azimuthal magnetic plasma fields~\cite{Khudik-POP_2016,arefiev2020PRE,gong2020PRE}. The laser pulse creates the quasi-static field structure if $\tau_l \gg \tau_e$. 

Laser pulses of ps duration, like the one available at OMEGA~EP and used in this paper (see \cref{appe: PIC} for parameters), are well-suited for DLA. Indeed, let us consider a plasma with a given electron density $n_e$. It is convenient to normalize $n_e$ to the classical critical/cutoff density $n_c = m_e \omega_0^2/4\pi e^2$, where $\omega_0$ is the laser frequency and $m_e$ and $e$ are the electron mass and charge. The plasma is transparent to the laser pulse regardless of its intensity if $n_e < n_c$. The characteristic electron response time is roughly equal to the period of plasma oscillations. Assuming that the bulk of plasma electrons is only weakly relativistic in the considered interaction, we find that $\tau_e~[\mbox{fs}] \approx 3.3 \sqrt{n_c/n_e}$ for a laser with a vacuum wavelength $\lambda_0=1.053~\micron$. Even in a weakly underdense plasma with $n_e/n_c = 0.02$ that we use in our simulations in this paper, the electron response time, $\tau_e \approx 23$~fs, is still very short compared to the ps laser pulse duration. The plasma density can even be lower, but such values are rarely used in practice because the charge of accelerated electron beam decreases with plasma density. We can conclude that a ps laser pulse is sufficiently long to establish the quasi-static field structure needed for DLA in a plasma with $n_e \gtrsim 0.01 n_c$. We therefore focus on this mechanism.

\begin{figure}[htb]
    \begin{center}
\includegraphics[width=1\columnwidth,clip]{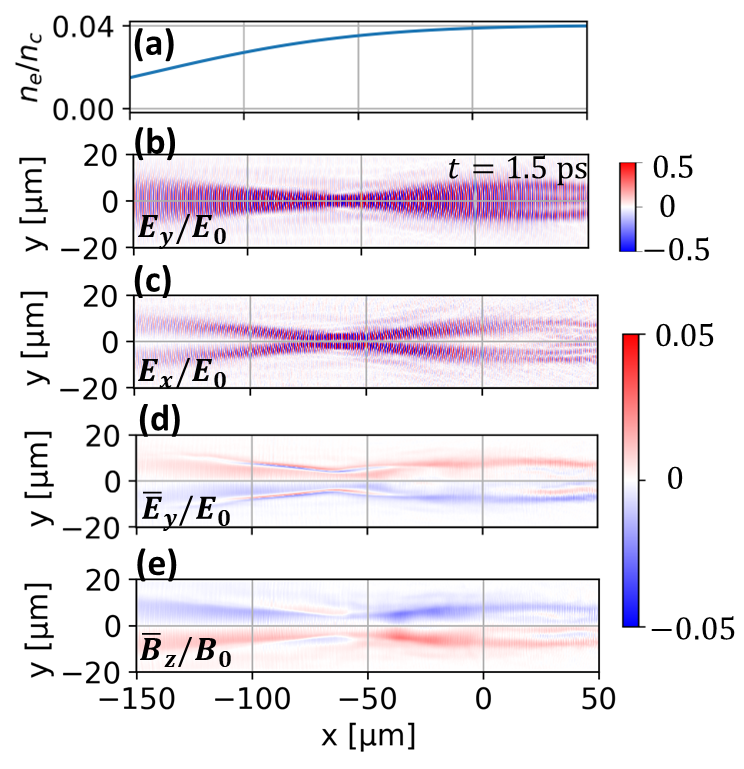}
\caption{\label{fig:field_DLA} The field structure during an interaction of a 700~fs long laser pulse with an underdense plasma whose initial profile is shown in (a). In the absence of the plasma, the laser peak intensity is $1.4\times 10^{20}$ W/cm$^2$ and the peak electric and magnetic fields are $E_0 = 10^{9}$~statvolt/cm and $B_0 = 10^9$~G. (b) \& (c) Transverse and longitudinal electric fields. (d) \& (e) Time-averaged transverse and longitudinal electric fields. The averaging is performed over five laser periods. }
    \end{center}
\end{figure}

An example of the laser-plasma interaction in the regime where the laser pulse duration greatly exceeds the characteristic electron response time is shown in \cref{fig:field_DLA}. The plots were obtained by performing a 2D PIC simulation (see \cref{appe: PIC} for details). The initial density profile is shown in \cref{fig:field_DLA}(a). Figures \ref{fig:field_DLA}(b) and \ref{fig:field_DLA}(c) show snapshots of transverse and longitudinal electric fields at $t = 1.5$~ps. The fields in these plots are primarily dominated by oscillating laser fields. However, time-averaging over 5 laser periods reveals that there are also quasi-static, i.e. slowly-varying, fields generated and sustained by the plasma. The electric field in \cref{fig:field_DLA}(d) is a result of charge separation created by transverse electron expulsion. The magnetic field in \cref{fig:field_DLA}(e) is a result of a longitudinal electron current created by a forward push from the laser.  

As already stated, the quasi-static radial electric and azimuthal magnetic plasma fields are critically important for effective DLA. During DLA, an electron gains its energy directly from the laser electric field. Typically, the energy comes from the transverse laser electric field. It is important to use laser intensities that make electrons relativistic. This is because the laser magnetic field effectively redirects the accumulated energy of relativistic electrons towards the longitudinal motion with the help of the $\bm{v} \times \bm{B}$ force. However, transverse electron expulsion from the laser beam and electron dephasing (slippage with respect to laser wavefronts) can terminate or limit the energy gain~\cite{arefiev2016beyond,arefiev2020PRE,gong2020PRE}. The quasi-static plasma fields provide transverse electron confinement within the laser beam by deflecting outward moving electrons towards the axis of the laser beam~\cite{Wang2020pop}. If the frequency of the transverse deflections matches the frequency of the laser field oscillations at the electron location, then there is a possibility for an enhanced electron energy gain due to a so-called betatron resonance~\cite{pukhov1999DLA}. Reference~[\onlinecite{Khudik-POP_2016}] provides a detailed analytical analysis of the energy gain. The enhanced energy gain has been observed in experiments~\cite{hussein2021njp} and numerical simulations~\cite{pukhov1999DLA,babjak.prl.2024}.  

In the context of x-ray and gamma-ray emission, there are several avenues for utilizing the electrons produced via DLA. One option is to send the electrons into a thick high-Z converter where they collide with atomic nuclei and emit energetic photons via bremsstrahlung~\cite{Edwards_APL_2002}. Another option is to leverage the same quasi-static radial electric and azimuthal magnetic plasma fields. This usually happens automatically, because the already discussed transverse deflections lead to synchrotron emission of electromagnetic radiation. We employ the term synchrotron emission to encompass the radiation emitted by a moving electron as it undergoes acceleration due to macroscopic fields. The synchrotron emission in laser-irradiated plasmas has been extensively studied for short laser pulses ($\sim 100$~fs) of ultra-high intensity ($> 10^{22}~\rm{W/cm^2}$). Such lasers can drive GG magnetic fields in a dense plasma. The fields have been shown to induce very efficient conversion of laser energy into gamma-ray radiation via synchrotron emission~\cite{Stark2016PRL,wang_power_PRA2020}. In general, the conversion efficiency increases with laser intensity, which explains why short laser pulses that are able to achieve very high peak intensity have received most of the attention. The synchrotron emission in plasmas irradiated by ps-long laser pulses remains relatively unexplored~\cite{Kneip_PRL_2008}. Our paper aims to fill this gap in knowledge by providing an in-depth analysis of x-ray emission by a finite-length plasma. 


\section{Estimates for synchrotron emission} \label{sec-estimation}

In \cref{sec-background}, we discussed that a ps-long laser pulse creates slowly evolving electric and magnetic fields in the plasma. The purpose of this section is to provide simple expressions for assessing the impact of these fields on the photon emission. We consider two limiting cases that are relevant to the emission in a finite-length plasma: a forward-moving ultrarelativistic electron and a backward-moving ultrarelativistic electron. At this stage, we simply treat the electron energy as being given without discussing electron acceleration.

In general, the radiation emitted by a moving electron as it undergoes acceleration due to macroscopic electric and magnetic fields ($\bm{E}$ and $\bm{B}$) is characterized by a single dimensionless parameter~\cite{jackson1999classical}
\begin{equation}
    \chi=\frac{\gamma}{B_{crit}} \sqrt{ \left( \bm{E} + \frac{1}{c} [ \bm{v} \times \bm{B} ] \right)^2 - \frac{1}{c^2} \left(\bm{E} \cdot \bm{v} \right)^2 }, \label{chi}
\end{equation}
where $\bm{v}$ is the electron velocity, $\gamma=1/\sqrt{1-v^2/c^2}$ is its relativistic factor, $c$ is the speed of light, and $B_{crit}=m_e^2 c^3 / (e \hbar) \approx 4.4 \times 10^{13}$~G is the critical magnetic field defined using the reduced Planck constant  $\hbar$. The emitted power is proportional to $\chi^2$. The spectrum of emitted photons has a peak at the energy equal to
\begin{equation}
\varepsilon_{\gamma}=1.5 \gamma \chi m_e c^2. \label{eq:ph_energy}
 \end{equation}
An ultrarelativistic electron emits primarily in the forward direction (along $\bm{v}$). In this case, the emitted power is concentrated within a cone with  a very small opening angle of $1/\gamma$. 

Motivated by the 2D PIC simulation shown in \cref{fig:field_DLA}, we consider an electron that experiences electric and magnetic fields, 
\begin{eqnarray}
    &&\bm{E}= E_x \bm{e}_x + E_y \bm{e}_y, \\
    &&\bm{B}=B_z \bm{e}_z,
\end{eqnarray}
that are a superposition of oscillating laser fields ($E^{laser}_x$, $E^{laser}_y$, and $B^{laser}_z$) and quasi-static plasma fields ($\overline{E}_y$ and $\overline{B}_z$):
\begin{eqnarray}
&&E_x= E^{laser}_x, \\
&&E_y=E^{laser}_y+ \overline{E}_y, \\
&&B_z=B^{laser}_z+ \overline{B}_z.
\end{eqnarray}
The laser propagates in the forward direction along the $x$-axis. In general, the electron has two velocity components: $\bm{v}= v_x \bm{e}_x+ v_y \bm{e}_y$. Therefore, the expression under the square root in \cref{chi} that we denote as
\begin{equation}
    {\cal{E}} = \left( \bm{E} + \frac{1}{c} [ \bm{v} \times \bm{B} ] \right)^2 - \frac{1}{c^2} \left(\bm{E} \cdot \bm{v} \right)^2 
\end{equation}
reduces to
\begin{eqnarray}
{\cal{E}}
&& = E_x^2 \left(1-\frac{v_x^2}{c^2}\right) + E_y^2 \left(1-\frac{v_y^2}{c^2}\right) + B_z^2 \frac{v^2}{c^2} \nonumber  \\
&& + \frac{2}{c} B_z (E_x v_y - E_y v_x) - 2 E_x E_y \frac{v_x v_y}{c^2}.  \label{eq: 8}
\end{eqnarray}
The relative orientation of the fields with respect to the direction of electron motion (the direction of $\bm{v}$) is clearly important in determining the value of ${\cal{E}}$ and, as a result, the value of $\chi$. 

It is instructive to first consider a simple case where there is no plasma and thus there are no plasma fields. If the laser is a vacuum plane wave, which implies that $B_z = E_y$, then we obtain from Eq.~(\ref{eq: 8}) that 
\begin{equation}
    {\cal{E}} = E_y^2 \left( 1 - \frac{v_x}{c} \right)^2.
\end{equation}
At fixed electron energy, $(1-v_x/c)$ has the smallest value in the case of forward motion. This reflects a well-known fact that forward-moving electrons are not efficient emitters of radiation in a plane electromagnetic wave. The expression $(1-v_x/c)$ has the largest value in the case of backward electron motion. This feature is often used to generate x-rays in a setup where a laser beam collides with an energetic electron beam~\cite{tsai.pop.2015, taphuoc.natphot.2012,vranic.prl.2014}. 

Our next step is to assess the impact of the quasi-static plasma fields. For simplicity, we only consider backward and forward moving electrons. It follows from Eq.~(\ref{eq: 8}) that for these electrons we have
\begin{equation}
{\cal{E}} = E_x^2 \left(1-\frac{v_x^2}{c^2}\right) + \left[ E_y - \frac{v_x}{c} B_z \right]^2,  \label{eq: 10}
\end{equation}
where it is taken into account that $v^2 = v_x^2$. The transverse fields of the laser are the dominant fields. Let us then start by retaining only these fields, with
\begin{equation}
{\cal{E}} \approx \left[ E_y^{laser} - \frac{v_x}{c} B_z^{laser} \right]^2  .  \label{eq: 11}
\end{equation}
This approximate expression is sufficient for the case of backward moving electrons. In the regime of interest, the plasma is significantly underdense, so that the amplitude of $B^{laser}_z$ is comparable to the amplitude of $E^{laser}_y$. We then set $E^{laser}_y \approx B^{laser}_z$ and find that ${\cal{E}}$ for backward moving ultrarelativistic electrons is simply given by
\begin{equation}
{\cal{E}}_{bwd} \approx 4(B_z^{laser})^2 .
\end{equation}
After substituting this expression into \cref{chi}, we find that the value of $\chi$ for a backward moving electron with $\gamma = \gamma_{bwd}$ is 
\begin{equation}
    \chi_{bwd} \approx 2\gamma_{bwd} \frac{|B^{laser}_z|}{B_{crit}}. \label{eq:bwd_chi}
\end{equation}
Therefore, as expected, the impact of the plasma fields on emission on backward-moving electrons is insignificant.

In the case of forward moving electrons, \cref{eq: 11} is no longer adequate for finding ${\cal{E}}$. Indeed, the terms on the right-hand side nearly cancel each other out for $E^{laser}_y \approx B^{laser}_z$. This means that the plasma fields, even though they are relatively weak, must be retained. We thus have
\begin{equation}
    E_y - \frac{v_x}{c} B_z = \left( E_y^{laser} - \frac{v_x}{c} B_z^{laser} \right) + \left[ \overline{E}_y - \frac{v_x}{c} \overline{B}_z \right].
\end{equation}
We again set $v_x \approx c$. To correctly determine the value of the expression inside the round brackets, we need to take into account the difference between $B^{laser}_z$ and $E^{laser}_y$. The laser produces a channel in the plasma that acts as a wave-guide. The laser beam can then be viewed as a wave-guide mode with  $E^{laser}_y = B^{laser}_z v_{ph} / c$, where $v_{ph}$ is the phase velocity of the mode. This velocity is superluminal ($v_{ph} > c$), making $B^{laser}_z$ lower than $E^{laser}_y$. We eliminate $E^{laser}_y$ using this relation and find that ${\cal{E}}$ for a forward moving ultrarelativistic electron with $\gamma = \gamma_{fwd}$ is given by
\begin{equation}
    {\cal{E}}_{fwd} \approx \left[ B^{laser}_z \left( \frac{v_{ph} - c}{c} \right) + \overline{E}_y - \overline{B}_z \right]^2.  
\end{equation}
The resulting value of $\chi$ is
\begin{equation} \label{eq:fwd_chi}
    \chi_{fwd} \approx \frac{\gamma_{fwd}}{B_{crit}} \left| B^{laser}_z \left( \frac{v_{ph} - c}{c} \right) + \overline{E}_y - \overline{B}_z \right|.
\end{equation}
This expression indicates that quasistatic plasma fields can greatly enhance $\chi_{fwd}$ and thus the photon emission by forward-moving electrons. \Cref{eq:fwd_chi} would acquire an additional $1/2 \gamma_{fwd}^2$ term inside the round brackets if we were to account for the difference between $v_x$ and $c$. This correction is relatively small and this is the reason why we set $v_x = c$. 

To conclude this section, we compare the energies of photons emitted by backward- and forward-moving electrons. We are particularly interested in the regime where the plasma fields enhance $\chi_{fwd}$, so we neglect the term that involves $B_z^{laser}$ in \cref{eq:fwd_chi}. Using the expression  given by \cref{eq:ph_energy} for the photon energy, we find that
\begin{equation}
    \frac{\varepsilon_{\gamma}^{fwd}}{\varepsilon_{\gamma}^{bwd}} \approx \frac{\gamma_{fwd}^2}{\gamma_{bwd}^2} \frac{\left| \overline{E}_y - \overline{B}_z \right|}{2|B^{laser}_z|} ,
\end{equation}
where $\varepsilon_{\gamma}^{fwd}$ is the energy of forward-moving photons emitted by forward-moving electrons and $\varepsilon_{\gamma}^{bwd}$ is the energy of backward-moving photons emitted by backward-moving electrons. Since the laser fields are the dominant fields, we can have $\varepsilon_{\gamma}^{bwd} \gtrsim \varepsilon_{\gamma}^{fwd}$ even if $\gamma^{bwd} \ll \gamma^{fwd}$. This aspect is important for understanding the simulation results shown in the next section.


\section{X-ray emission at density down-ramp} \label{sec: back_emission}

\begin{figure*}[!htb]
    \begin{center}
    \includegraphics[width=0.7\textwidth]{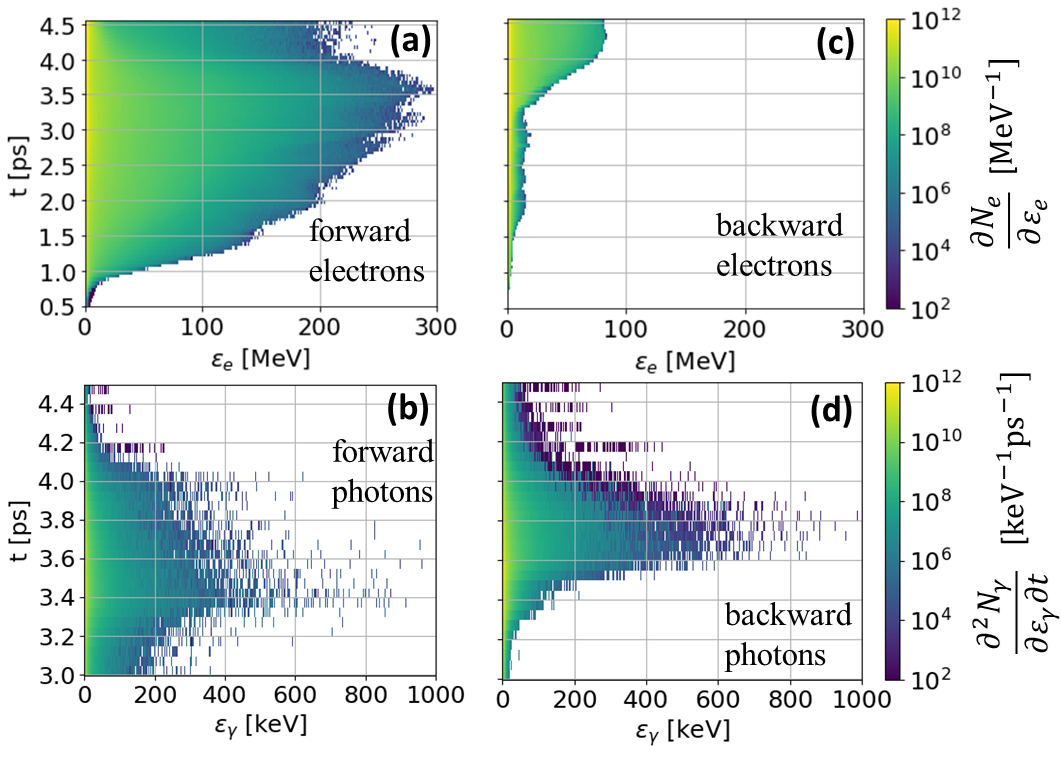}
    \caption{\label{fig:time_evolution} Time-evolution of the electron energy spectra (left column) and  snapshots of the energy distribution of emitted photons (right column). (a)\&(c) Electron energy spectra for forward- and backward-moving electrons. The spectra are recorded every 50~fs. The color-coding gives the number of electrons per MeV. (b)\&(d) Energy spectra of forward and backward emitted photons. The spectra are calculated by accumulating photon emissions over 50~fs intervals. The color-coding gives the number of photons per keV per ps. }
    \end{center}
\end{figure*}

In this section, we examine the photon emission in the 2D PIC simulation shown in \cref{fig:field_DLA}. The choice of the simulation setup and parameters is motivated by DLA experiments that can be performed using the OMEGA EP laser system (e.g. see Ref.~[\onlinecite{tang.njp.2024}]).

In the simulation, a high-intensity 700~fs laser beam irradiates an under-dense helium plasma of a finite length. The beam is linearly polarized, with a peak intensity of $I_0 = 1.4\times 10^{20}$ $\rm{W/cm^2}$. The laser wavelength is $\lambda_0=1.053~\micron$, so the selected $I_0$ corresponds to $a_0 \equiv |e|E_0/m_e c\omega_0 = 10$, where $E_0$ is the peak amplitude of the laser electric field. Additional details regarding the simulation setup, including laser beam and target parameters are given in \cref{appe: PIC}. An important feature of the selected parameters is that the laser pulse is able to go through the plasma before becoming fully depleted. The plasma density is nonuniform, with a super-Gaussian profile shown in \cref{fig:ne_profile} in \cref{appe: PIC}. We are particularly interested in the laser-plasma interaction at the density down-ramp, as this interaction has received relatively little attention.

In agreement with our expectations, the laser beam generates a population of energetic forward-directed electrons via the DLA mechanism. \Cref{fig:time_evolution}(a) shows the time evolution of the energy distribution for these electrons. The cut off electron energy starts to increase at around $t = 0.8$~ps. It continues to grow until $t=3.5$~ps, reaching around 290~MeV. The laser enters the density down-ramp slightly before the cutoff energy reaches its peak value. 

We want to point that $N_e$ in \cref{fig:time_evolution}(a) is the total number of forward-moving electrons in a plasma that is $8~\micron$ thick along the third dimension. By default, a 2D EPOCH simulation outputs a spectrum where the number of particles is given per meter along the third dimension. The plotted spectrum is the default spectrum multiplied by $\Delta = 8~\micron$, which is the diameter of the laser focal spot. We chose this way of presenting our spectrum because then it gives a meaningful number of particles. We present all our spectra in this paper this way. 

The emission by the forward-moving electrons is shown in \cref{fig:time_evolution}(b). In the simulation,  photons are emitted as individual particles in the direction of the electron momentum. The emission is calculated during the PIC simulation using the Monte Carlo algorithm described in Refs.~[\onlinecite{ridgers.jcp.2014}] and [\onlinecite{gonoskov.pre.2015}]. The algorithm uses the value of $\chi$ for each electron to determine the photon energy. The electron experiences recoil as a result of the emission. \Cref{fig:time_evolution}(b) shows the energy spectrum of the photons  emitted at a given time $t$ within a 50~fs window. We see from \cref{fig:time_evolution}(b) that the emitted photon energies increase and peak at around $t=3.5$~ps. Not surprisingly, this trend matches the trend observed for the electron energies.  

The emission of the energetic forward photons is primarily caused by quasi-static plasma fields. This can be confirmed using \cref{eq:fwd_chi} for $\chi_{fwd}$ derived in \cref{sec-estimation} and the fields experienced by energetic electrons at $t \approx 3.5$~ps. Based on the simulation results, we have $|\overline{E}_y| \approx 0.02 E_0$ and $|\overline{B}_z| \approx 0.05 B_0$. These fields have opposite signs, which means that their contributions to $\chi_{fwd}$ add up. The laser magnetic field is roughly $|B^{laser}_z| \approx 0.5 B_0$. By tracking laser wavefronts we found that $v_{ph}/c = 1.0028$. Therefore, the magnitude of the term that involves $B^{laser}_z$ in the expression for $\chi_{fwd}$ is roughly $1.4 \times 10^{-3} B_0$. The amplitude of the contribution from $\overline{E}_y - \overline{B}_z$ is greater by a factor of 50. We thus have 
\begin{equation}
    \chi_{fwd} \approx \gamma_{fwd} \left. \left| \overline{E}_y - \overline{B}_z \right| \right/ B_{crit}.
\end{equation}
For an electron with $\gamma_{fwd} \approx 400$ this expression yields $\chi_{fwd} \approx 6.4 \times 10^{-4}$. We find from \cref{eq:ph_energy} that the spectrum of photons emitted by this electron has a peak at $
\varepsilon_{\gamma} \approx 200$~keV, which is roughly the photon energy range observed in our simulation.

\begin{figure}[htb]
    \begin{center}
    \includegraphics[width=1\columnwidth,clip]{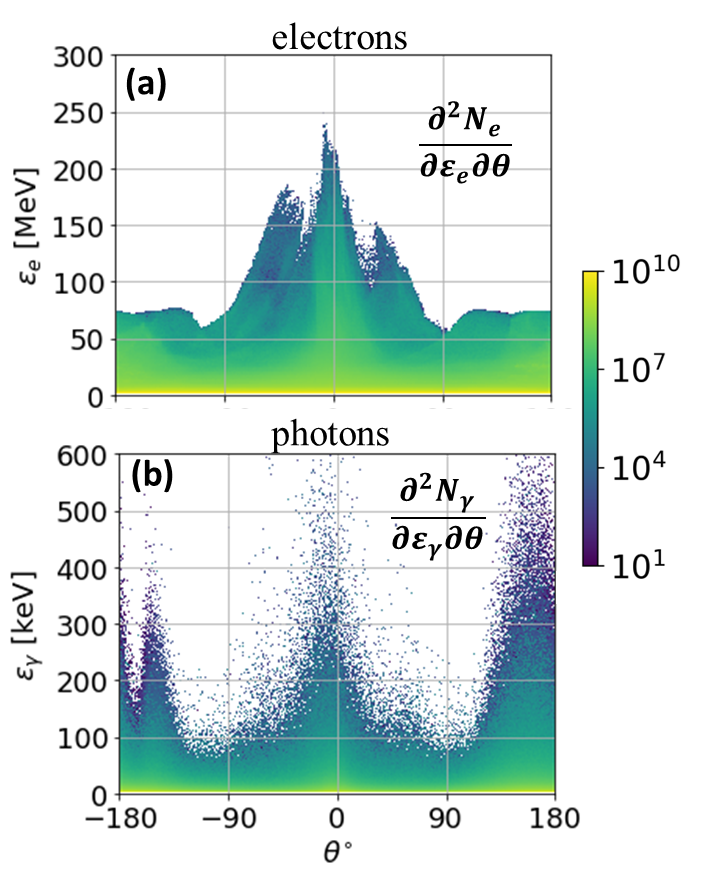}
    \caption{\label{fig:field_spectrum} Angle-resolved electron and photon spectra during the propagation of the 700~fs laser pulse along the density down-ramp ($t = 4$~ps). For electrons, the color is the number of electrons per MeV per degree, i.e.  $1/(\rm{MeV}\,\rm{deg}^{\circ})$. For photons, the color is the number of photons per keV per degree, i.e. $1/(\rm{keV}\,\rm{deg}^{\circ})$.}
    \end{center}
\end{figure}

The angle-resolved spectrum of the photons emitted by $t = 4$~ps is shown in \cref{fig:field_spectrum}(b),  where $\theta$ is the conventional polar angle measured with respect to the positive direction of the $x$ axis. As expected, there is a forward-directed beam of energetic photons in the  $\gamma$-ray energy range. These are the photons emitted by the forward-moving electrons generated via DLA. The spectrum also has an unexpected feature: in addition to the forward photons, there is a significant population of equally energetic photons that have been emitted backwards. These photons are necessarily emitted by energetic backward-moving electrons that are generated by a mechanism that is different from DLA.    

Figures~\ref{fig:time_evolution}(c) and \ref{fig:time_evolution}(d) show the time evolution of the energy spectra of backward electrons and backward photons. In contrast to the forward electrons, energetic backward electrons only appear at $t > 3.25$~ps when the laser pulse starts propagating along the density down-ramp. As seen in \cref{fig:time_evolution}(d), the emission of energetic backward photons is directly correlated with the emergence of the energetic backward electron population. \Cref{fig:field_spectrum}(a) provides an angle-resolved spectrum of the electrons in the entire simulation domain at $t = 4$~ps. Here $\theta$ is angle between the electron momentum vector and the $x$-axis. In contrast to the forward-moving electrons, backward-moving electrons lack a directed energetic peak. Their maximum energy is about 70~MeV over a broad range of angles.

The emission of the backward photons is caused by the laser pulse itself. This is because the population of the energetic backward electrons overlaps with the laser pulse. The parameter $\chi$ for the electrons that move directly backward is given by \cref{eq:bwd_chi}. We take $|B^{laser}_z| \approx 0.3 B_0$, so that for an electron with $\gamma_{bwd} \approx 140$ we get $\chi_{bwd} \approx 1.9 \times 10^{-3}$. It then follows from \cref{eq:ph_energy} that the spectrum of photons emitted by this electron has a peak at $
\varepsilon_{\gamma} \approx 206$~keV. This energy is comparable to what is shown in \cref{fig:field_spectrum}(a) for backward-emitted photons.

\begin{figure*}[!htb]
    \begin{center}
    \includegraphics[width=0.7\textwidth]{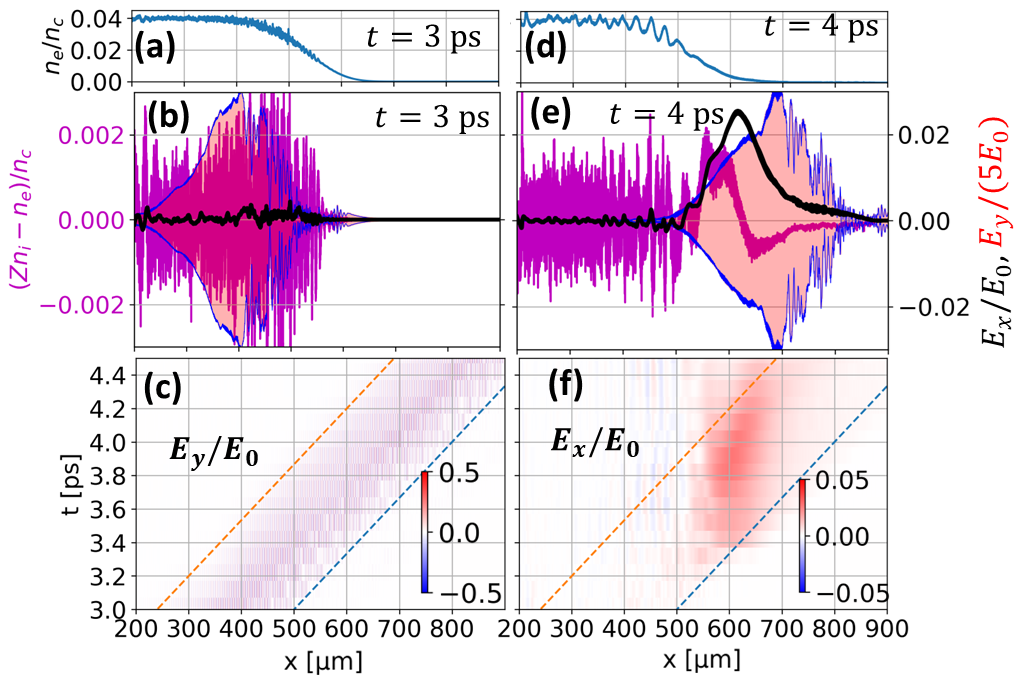} \caption{\label{fig:field_evolution} Impact of the density down-ramp. (a)\&(d) Electron density at $t = 3$~ps and at $t = 4$~ps. (b)\&(e) Normalized charge density and transverse and longitudinal electric fields at $t = 3$~ps and at $t = 4$~ps. The solid blue curve shows the laser pulse envelope and the transparent red color shows the actual transverse electric field $E_y$. Due to the small wavelength of the oscillations, individual oscillations are not discernible, and the plot appears as a solidly shaded region. (c)\&(f) Time-evolution of $E_y$ and $E_x$.  All the quantities are averaged over a transverse slice with $|y| \leq 30$ $\micron$.   }
    \end{center}
\end{figure*}

There are two factors that influence the angular dependence of the backward photon spectrum in \cref{fig:field_spectrum}(a). Recall that \cref{eq:bwd_chi} for $\chi_{bwd}$ was derived assuming that the electron is moving directly backwards. If the electron is moving at an angle with respect to the $x$-axis, then its $\chi$ is lower than $\chi_{bwd}$. This is the reason why the photon spectrum peaks in the backward direction even though the electron spectrum is pretty flat. Laser filamentation is another factor that impacts the emission shown in \cref{fig:field_spectrum}(a). We found that the laser beam becomes very distorted at the density down-ramp. The distortion can be viewed as an additional tilt of wavefronts that fluctuates across the beam. The implication is that the backward direction with $|\theta| = 180^{\circ}$ is no longer the direction that universally has the highest value of $\chi$. As a result, there is an additional peak at $\theta \approx -150^{\circ}$ in the considered simulation. It is shown in \cref{appe: multiple_run} that this pattern changes when we change the random seed that is used to initialize the plasma in the PIC simulation. However, the general trend of enhanced backward-directed emission at the density down-ramp is robust.


\section{Generation of backward-moving electrons} \label{sec: backward electrons}

In the previous section, we showed that the ps-long laser pulse used in our simulation induces backward-directed $\gamma$-ray emission at the density down-ramp. The photons are emitted by energetic backward-moving electrons. In this section we detail the mechanism that generates these electrons.

To gain insight into electron dynamics, we examined the time evolution of transverse ($E_y$) and longitudinal ($E_x$) electric fields. Figures~\ref{fig:field_evolution}(c) and \ref{fig:field_evolution}(f) show the time evolution of $E_y$ and $E_x$ along $x$ after the laser pulse enters the density down-ramp. Dashed lines are added to both plots to roughly mark the extent of the laser pulse at each $t$. It is apparent from \cref{fig:field_evolution}(f) that the laser pulse generates a strong longitudinal plasma electric field at the density down-ramp by pushing plasma electrons forward. The field is slowly evolving compared the laser oscillations and its longitudinal scale greatly exceeds the laser wavelength.  The amplitude of this field is roughly $3\%$ of the amplitude of the transverse laser electric field. The amplitude of the plasma electric field is much higher than the amplitude of the oscillating longitudinal laser electric field, so the field in \cref{fig:field_evolution}(f) is essentially $\overline{E}_x$.

\begin{figure*}[!t]
    \begin{center}
    \includegraphics[width=0.8\textwidth]{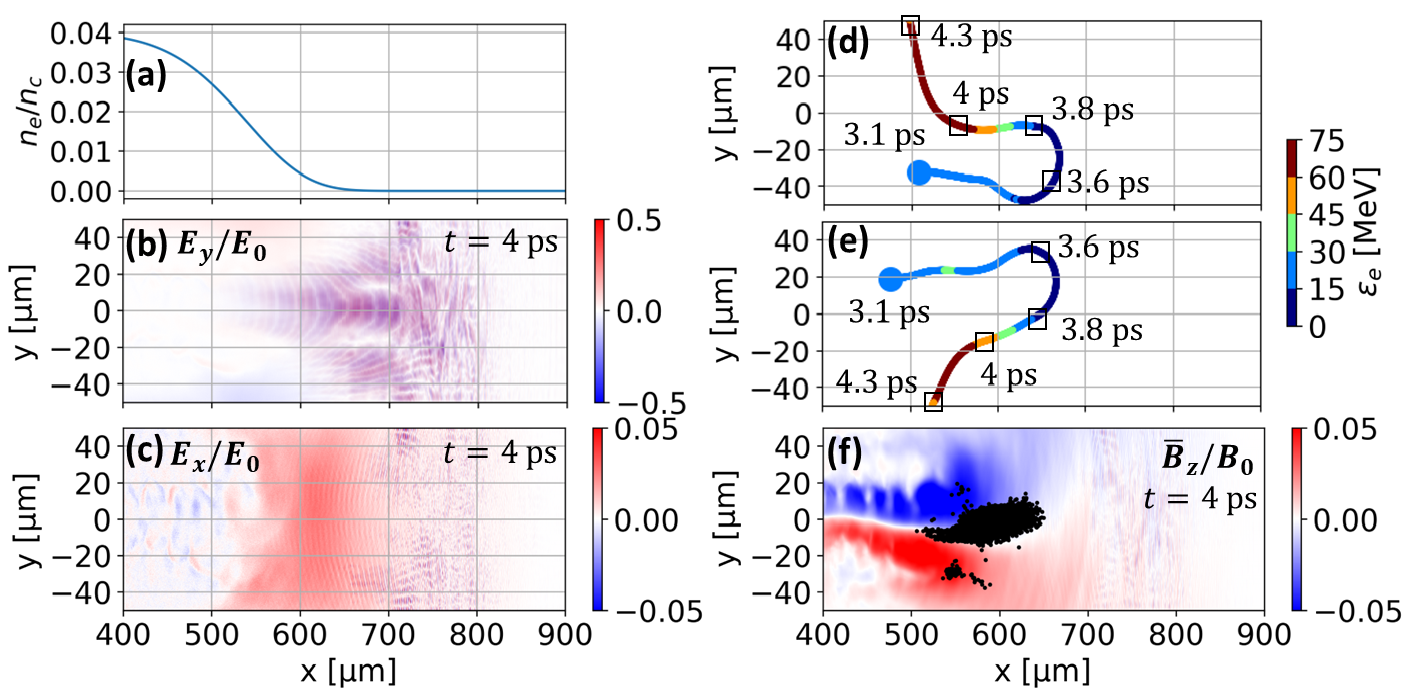}
    \caption{\label{fig:e_dyanmics} (a) Initial electron density profile. (b) Transverse electric field $E_y$ at $t = 4$~ps. (c) Longitudinal electric field $E_x$ at $t = 4$~ps.
    (d)\&(e) Two electron trajectories for $3.1~\mbox{ps}\leq t \leq 4~\mbox{ps}$. The color is the energy of the electrons. (f) Time-averaged plasma magnetic field at $t = 4$~ps. The black dots are the locations of backward photon emissions with $\varepsilon_{\gamma}>100$ keV and $|\theta - 180^{\circ}| \leq 10^{\circ}$ from the start of the simulation ($t \leq 4$~ps). }
    \end{center}
\end{figure*}

The longitudinal field at the down-ramp is a charge-separation electric field. To substantiate this statement, Figs.~\ref{fig:field_evolution}(a)~-~(e) provide additional information about the laser pulse propagation before and after it enters the density down-ramp ($t = 3$~ps and $t = 4$~ps). In Figs.~\ref{fig:field_evolution}(b) and (e), the red curve is $E_y$, the black curve is $E_x$, and the magenta curve is the charge density. All three quantities are averaged over a transverse slice with $|y|\leq 30~\micron$ to distinguish a global trend from local fluctuations. Note that the charge density is normalized to $|e|n_c$, so the plotted quantity is $(Z n_i-n_e)/n_c$, where $n_e$ is the electron density, $n_i$ is the ion density, and $Z=2$ is the ion charge number. At $t=3$~ps, the laser pulse is in a region without a significant density gradient, as seen from comparing  \cref{fig:field_evolution}(a) and \cref{fig:field_evolution}(b). The charge density generated by the laser in this region rapidly oscillates along $x$, so that the longitudinal electric field $E_x$ is very weak. At $t=4$~ps, the laser pulse is at the bottom of the density down-ramp, as seen from comparing  \cref{fig:field_evolution}(d) and \cref{fig:field_evolution}(e). By pushing electrons forward, the pulse generates charge separation on a large spatial scale. The resulting charge density creates and sustains a positive $E_x$, with the direction of the field opposite to the direction of the density gradient.

The extent and the magnitude of $E_x$ are such that this field can turn around moderate energy DLA electrons entering the density down-ramp. It is convenient to write down the work done on an electron by $E_x$ with a longitudinal extent $l$ as 
\begin{equation}
    \frac{|e| E_x l}{m_e c^2} = 2 \pi a_0 \frac{l}{\lambda_0} \frac{E_x}{E_0},  
\end{equation}
where $\lambda_0 \approx 1~\micron$ is the laser wavelength in vacuum. Taking $E_x \approx 2 \times 10^{-2} E_0$ and $l \approx 100 \lambda_0$, we find that $|e| E_x l / m_e c^2 \approx 120$. This means that electrons with energy $\varepsilon_e \lesssim 60$~MeV can be stopped by a static $E_x$ with the considered amplitude and extent. Once the electrons stop, they can be re-accelerated backwards by the same longitudinal field. As a result, the plasma $E_x$ produces a population of energetic backward-moving electrons.

To confirm electron re-acceleration, we have tracked multiple electrons in our simulation. The electrons were selected by randomly picking several backward-moving electrons at $t = 4$~ps with kinetic energies above 50~MeV. As shown in Figs.~\ref{fig:e_dyanmics}(a)~-~(c), the laser pulse is already at the bottom of the density ramp at $t = 4$~ps and it has generated a strong large-scale $E_x$. Two representative trajectories are shown in \cref{fig:e_dyanmics}(d) and \cref{fig:e_dyanmics}(e), with the color indicating kinetic electron energy. Both of these electrons enter the density down-ramp moving forward. At $t = 3.1$~ps their kinetic energy is about 20~MeV. After traveling roughly $100~\micron$, they lose most of their kinetic energy due to the work performed by $E_x$. Once they turn around, they indeed start re-accelerating and gaining energy, confirming the qualitative picture given earlier. 

The energy gain during the re-acceleration process is aided by the evolution of $E_x$. The forward longitudinal electric field $E_x$ builds up on a time scale comparable to the time needed for the electrons to turn around. As a result, backward-moving electrons experience a much stronger field. This aspect can be deduced from the evolution of the electron energy shown in \cref{fig:e_dyanmics}(d) and \cref{fig:e_dyanmics}(e). Indeed, by the time the two electrons come back to the top of the density ramp, their kinetic energy is noticeably higher ($>60$~MeV) than the energy they had when they entered the down-ramp ($\sim 20$~MeV). This means that $E_x$ not only turns moderate energy DLA electrons around, but it also makes them more energetic.

\begin{figure*}[!htb]
    \begin{center}
    \includegraphics[width=0.7\textwidth]{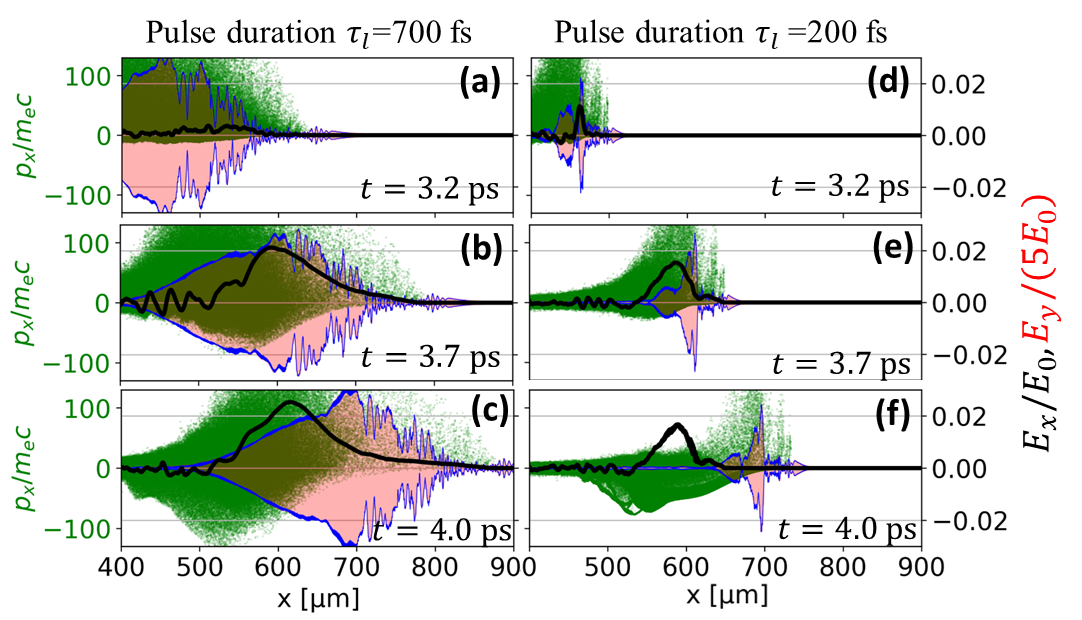}
    \caption{\label{fig:long_short} A time sequence of snapshots for the 700~fs and 200~fs laser pulses. 
    The green dots show the electron phase space. The solid blue curve shows the laser pulse envelope and the transparent red color shows the actual transverse electric field $E_y$. Due to the small wavelength of the oscillations, individual oscillations are not discernible, and the plot appears as a solidly shaded region. The black curves show the time-averaged (over 5 laser periods) longitudinal electric field $\overline{E}_x$. Both fields are averaged over a transverse slice with $|y|\leq 30$ $\micron$.  }
    \end{center}
\end{figure*}

The turning around of the electrons is facilitated by the plasma magnetic field shown in \cref{fig:e_dyanmics}(f). To illustrate this, we perform several simple estimates of the electron gyroradius,
\begin{equation}
    \frac{r_e}{\lambda_0} \approx \frac{\gamma}{2 \pi a_0} \frac{B_0}{\overline{B}_z},
\end{equation}
using $\overline{B}_z=0.01 B_0$, which is the characteristic field strength experienced by the electrons. At the top of the down-ramp ($t = 3.1$~ps), $\varepsilon_e \approx 20$~MeV and $\gamma \approx 40$, so the electron gyroradius is $r_e \approx 60 \lambda_0 \approx 60~\micron$. This radius greatly exceeds the width of the magnetic filament in \cref{fig:e_dyanmics}(f). As a result, a forward-moving electron is not able to turn around purely due to the plasma magnetic field. Instead, the magnetic field, whose sign changes as the electron crosses the axis of the filament, deflects the electron forward~\cite{gong2020PRE}. While the electrons slow down under the influence of $E_x$, their gyroradius decreases linearly with $\gamma$. At $\varepsilon_e \approx 2$~MeV, the gyroradius becomes sufficiently small, $r_e \approx 6 \lambda_0 \approx 6~\micron$, for the electron to turn around. As a result, the electrons transition to being backward-moving without coming to a complete stop. The turn-around time of roughly $\pi r_e/c \approx 62$~fs is consistent with the trajectories shown in Figs.~\ref{fig:e_dyanmics}(d) and \ref{fig:e_dyanmics}(e). 

After turning around, both electrons begin gaining energy, which causes $r_e$ to increase. This increase is sufficiently rapid to prevent electrons from quickly turning around one more time. The electrons then remain backward-moving, while $\overline{B}_z$ gradually deflects them away from the axis of the magnetic filament. This explains the change in electron trajectories between $t = 3.8$~ps and $t = 4.3$~ps in Figs.~\ref{fig:e_dyanmics}(d) and \ref{fig:e_dyanmics}(e).

We have so far discussed the electron dynamics along the density down-ramp while focusing on plasma electric and magnetic fields. These are the fields that are primarily responsible for the generation of backward-moving energetic electrons. The laser fields are also present, but they are rapidly oscillating. As a result, their impact on the global shape of the trajectory for backward-moving electrons is minor.

The roles of the laser and plasma fields become reversed when considering the photon emission by the backward-moving electrons. The laser fields are much stronger than the plasma field, so they can induce a much stronger acceleration. As discussed in \cref{sec-estimation}, the amplitude of the laser field sets the value of $\chi$ for backward-moving electrons and thus it is the laser field that primarily determines the emission process. The value of $\chi_{bwd}$ given by \cref{eq:bwd_chi} depends also on $\gamma$, so the electrons need to gain energy before the emission can become significant. In our case, the laser pulse is sufficiently long for the turned around electrons to gain high energy before leaving the pulse. \Cref{fig:e_dyanmics}(f) shows the locations where backward-directed photons with  $\varepsilon_{\gamma}>100$~keV were emitted by backward-moving electrons with $|\theta-180^{\circ}|<10^{\circ}$. The emissions are accumulated from the start of the simulation ($t\leq 4$~ps). The emissions are clustered in a region with a very weak plasma magnetic field, which confirms that they are caused by the laser rather than plasma fields. An important take-away point is that the need for the energetic electrons to overlap with the laser pulse imposes a  constraint on the pulse duration.

\section{Impact of pulse duration on backward emission} \label{sec: pulse duration}

In \cref{sec: backward electrons}, we showed that the longitudinal plasma field $E_x$ generated at the density down-ramp can turn some of the DLA electrons around and accelerate them backward. The fields of the laser cause ultrarelativistic backward-moving electrons to emit energetic backward photons. The emission strongly depends on the electron energy, but the energy gain process is not instantaneous. The electrons gain their energy from $E_x$, so it is conceivable for the electrons to reach their peak energy after they have already gone past the laser pulse. This is why the emission of backward-directed photons is sensitive to the duration of the laser pulse. In this section, we illustrate this sensitivity by comparing the emission induced by two laser pulses of different durations: the original 700~fs pulse and a shorter 200~fs pulse.

To help us compare the backward electron acceleration, \cref{fig:long_short} provides a time sequence of snapshots for both pulses. The sequence starts at $t = 3.2$~ps when the pulses enter the density down-ramp. The sequence ends at $t = 4$~ps when the two pulses reach the bottom of the ramp. To help us correlate the electron dynamics with the laser pulse propagation, each panel of the sequence shows the electron phase space (green), instantaneous transverse field $E_y$ (red), and time-average longitudinal field $\overline{E}_x$ (black). The fields are additionally averaged over a transverse slice with $|y|\leq 30$ $\micron$ to reduce the role of fluctuations. 

Prior to their descent pulses generate only forward-moving energetic electrons. The snapshots in \cref{fig:long_short}(a) and \cref{fig:long_short}(d) show that the situation remains relatively unchanged at the start of the descent. This is because both pulses have not yet had the opportunity to build up a long-scale slowly varying electric field $E_x$. The shorter pulse has a steeper intensity increase at the leading edge. This steep leading edge expels electrons more effectively, creating a spike of $E_x$ that is visible in \cref{fig:long_short}(d). This spike is however not stationary. It quickly moves forward with the laser pulse, which prevents the plasma $E_x$ from accelerating electrons backward to ultra-relativistic energies.

During their descent, both pulses generate a long-scale $E_x$ seen in Figs.~\ref{fig:long_short}(b) and \ref{fig:long_short}(e). This field can turn around some of the DLA electrons and accelerate them in the backward direction, as detailed in \cref{sec: backward electrons}. The span of $E_x$ generated by the shorter pulse is somewhat shorter, but what is even more striking is that this field peaks behind the laser pulse. A consequence of this is that there are no energetic electrons to collide with the laser field in the case of a shorter pulse.

As discussed in \cref{sec: backward electrons}, $E_x$ builds up on a time-scale comparable to the electron slow-down time. The time-evolution of $E_x$ causes the electrons that are re-accelerated backward to gain more energy than what they had moving forward. This is why the energy of backward electrons is higher in \cref{fig:long_short}(c) compared to \cref{fig:long_short}(b), and in \cref{fig:long_short}(f) compared to \cref{fig:long_short}(e). The longer pulse regime benefits from this phenomenon, whereas the shorter pulse regime does not, as the energy increase occurs well behind the laser pulse.

\begin{figure}[htb]
    \begin{center}
    \includegraphics[width=1\columnwidth,clip]{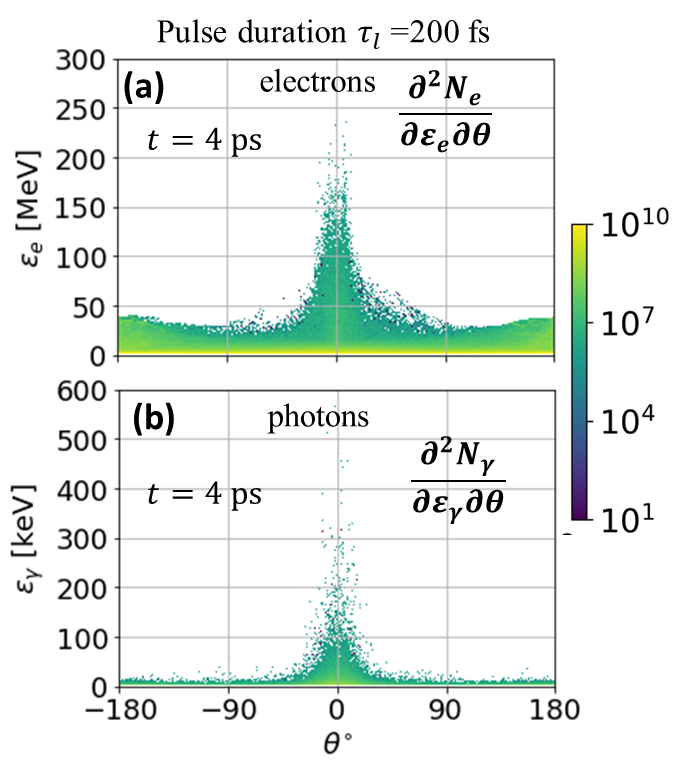}
    \caption{\label{fig:short_spectrum} Angle-resolved electron and photon spectra during the propagation of the 200~fs laser pulse along the density down-ramp ($t = 4$~ps). For electrons, the color is the number of electrons per MeV per degree, i.e.  $1/(\rm{MeV}\,\rm{deg}^{\circ})$. For photons, the color is the number of photons per keV per degree, i.e. $1/(\rm{keV}\,\rm{deg}^{\circ})$.}
    \end{center}
\end{figure}

\Cref{fig:short_spectrum} shows  the electron and photon spectra at $t = 4$~ps for the shorter pulse. The format is similar to that of \cref{fig:field_spectrum} to facilitate a comparison between the two pulses. The key features of the electron spectrum in \cref{fig:short_spectrum}(a) are similar to those in \cref{fig:field_spectrum}(a). There is a highly energetic population of electrons that is directed forward. These are the electrons generated via DLA. There is also a population of moderately energetic backward-moving electrons. These are the electrons generated by $E_x$ at the density down-ramp. The cutoff energy for backward electrons reaches 45~MeV, which is on par with what we see in \cref{fig:field_spectrum}(a) for the longer pulse. In contrast to the electron spectra, the photon spectra look qualitatively different. The photon spectrum in \cref{fig:short_spectrum}(b) has only a forward-directed peak. There is no emission of energetic photons in the backward direction even though the shorter laser pulse generates a population of moderately energetic backward-moving electrons. The angle-resolved photon spectrum confirms that the emission process of energetic backward-directed photons is sensitive to the laser pulse duration. The takeaway point of this section then is that backward emission at the density down-ramp requires a sufficiently long laser pulse.


\begin{figure*}
    \begin{center}
    \includegraphics[width=1\textwidth]{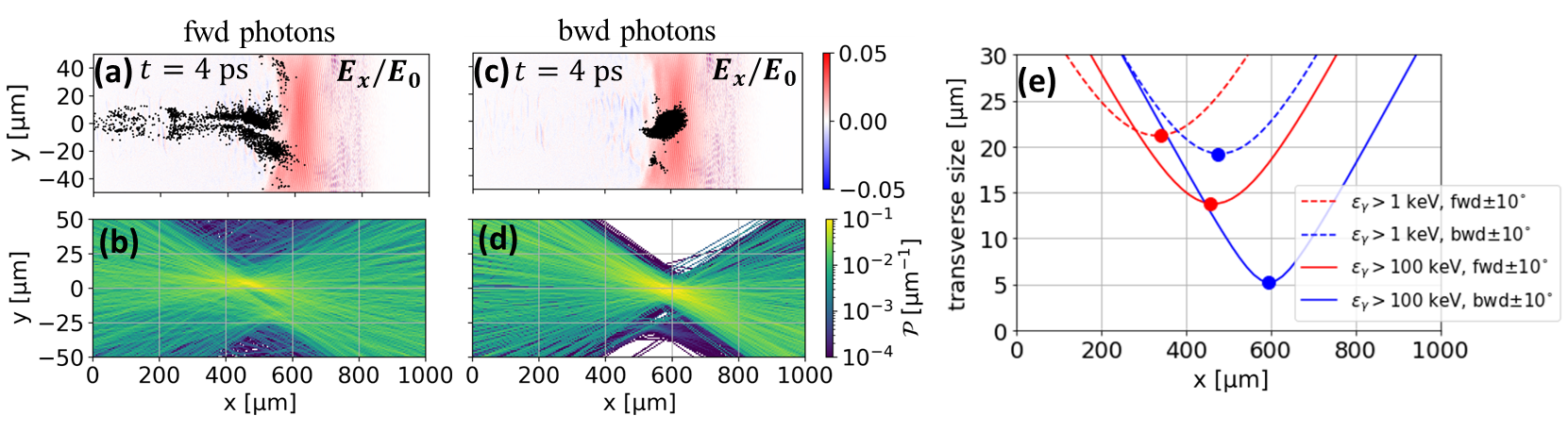}
    \caption{\label{fig:photon_source} (a)\&(c) Locations where forward- and backward-directed photons with $\varepsilon_{\gamma}>100$~keV were emitted over 4~ps ($t\leq 4$~ps). The background color is $E_x$. (b)\&(d) Normalized transverse distribution of rays, $\cal{P}$, calculated based on the photon emission data (see main text for details). (e)  Transverse size of the ray bundles, as those shown in (b) and (d), for different groups of photons. The round markers show what we call the source size.
}
    \end{center}
\end{figure*}

\section{Features of the emitted backward and forward photon beams} \label{sec: source size}

We have shown that the backward emission occurs at the density down-ramp and that this emission only takes place if the laser pulse is sufficiently long. To compare this emission with the better known forward emission, we calculate several key characteristics of the backward- and forward-directed photon beams generated in the plasma irradiated by the 700~fs laser pulse.

\Cref{fig:photon_source} shows the locations where photons with $\varepsilon_{\gamma}>100$~keV are emitted during 4~ps of the simulation. The black dots in \cref{fig:photon_source}(a) are for forward-directed photons with $|\theta|<10^{\circ}$, whereas the black dots in \cref{fig:photon_source}(c) are for backward-directed photons with $|\theta-180^{\circ}|<10^{\circ}$. The color in the background shows $E_x$ at $t = 4$~ps, with the extended red region roughly matching the location of the density down-ramp. As seen in \cref{fig:photon_source}(a), the forward-directed photons are emitted inside the bulk of the plasma over a $600~\micron$ interval in the longitudinal direction. The emission becomes more intense with the increase of $x$ because of the electron energy increases during the electron motion with the laser pulse [see \cref{fig:time_evolution}(a) for the time-evolution of the electron spectrum]. In contrast to the forward-directed photons, the backward-directed photons are exclusively emitted at the density down-ramp. As a result, their emission is localized within just a $100~\micron$ interval along $x$.

Once emitted, each $\gamma$-ray photon moves essentially in a straight line due to the low density of the plasma. Far away from the plasma, these photons are perceived as a bundle of diverging rays emitted by a photon source with a finite transverse size. The perceived source size is determined by the direction of the rays and, as a result, it might be different from the spatial extent of the actual region where the photons were emitted. The size of the source is important for imaging applications and thus it serves as an important metric for laser-driven photon emission.

To calculate the source size, we use the emission data provided by the EPOCH code. The output data contains the location and the direction of the emission for each emitted macro-particle representing photons. We first draw a ray through each emission location in the direction of the emission. We then flip the direction and draw another ray in the opposite direction. This step is important because the location of the source might be behind the actual emission location. The described procedure yields a bundle of rays (lines), where each ray has a weight set by the weight of the corresponding macro-particle. The weight determines how many real photons each macro-particle represents. The transverse distribution $\cal{P}$ of the rays for each $x$ position calculated using their weights is shown in \cref{fig:photon_source}(b) for the forward-emitted photons and in \cref{fig:photon_source}(d) for the backward-emitted photons. The distribution is normalized such that $\int_{-\infty}^{+\infty} {\cal{P}}(y) dy = 1$, where $y$ is measured in $\micron$. We define the transverse size of a bundle of rays at a given $x$ as the standard deviation of the transverse distribution. \Cref{fig:photon_source}(e) shows the dependence of the transverse size on $x$ for different groups of photons. We define the source size for a given group of photons as the smallest transverse size (shown with round markers).  

Our results presented in \cref{fig:photon_source}(e) highlight two trends. We found that the source size for forward- and backward-directed photons decreases as we increase the lower photon energy cut-off. The results are shown for photons with $\varepsilon_{\gamma} > 1$~keV and for photons with $\varepsilon_{\gamma} > 100$~keV. We also found that the source size for the backward-directed photons is smaller than that for the forward-directed photons. This feature is particularly pronounced at $\varepsilon_{\gamma} > 100$~keV, with the source size of $5~\micron$ for the backward photons being three times smaller than the source size for the forward photons. In the considered regime, the source size is most likely influenced by the longitudinal extent of the emission region discussed earlier in this section. As seen in \cref{fig:photon_source}(c), the backward emission is concentrated only within the density down-ramp. In contrast to this, the forward emission seen in \cref{fig:photon_source}(a) is spread over a significant part of the plasma.

Another two important characteristics of the emitted photon beams are their collimation and pointing. To assess these features for photons with $\varepsilon_{\gamma} > 100$~keV, we computed the mean and the standard deviation of $\theta$ for forward and backward photons in this energy range. We found that for the forward-directed photons the average angle is $|\theta_0^{fwd}| \approx 1.5^{\circ}$ and the standard deviation is $\Delta \theta^{fwd} \approx 5.2^{\circ}$. These quantities specify the pointing direction and collimation of the beam. For the backward-directed photons, it is convenient to measure the angle with respect to the backward direction, which effectively means an offset by $180^{\circ}$ for the conventional angle. Under this definition, the average angle is $|\theta_0^{bwd}| \approx 3.5^{\circ}$ and the standard deviation is $\Delta \theta^{bwd} \approx 4.8^{\circ}$. The pointing angles $|\theta_0^{fwd}|$ and $|\theta_0^{bwd}|$ change from run to run. However, their values are smaller than the corresponding standard deviations given by $\Delta \theta^{fwd}$ and $\Delta \theta^{bwd}$. This means that the photons beams remain well-directed forward or backward despite the inherent fluctuations caused by the plasma.

\begin{figure}[htb]
    \begin{center}
    \includegraphics[width=1\columnwidth,clip]{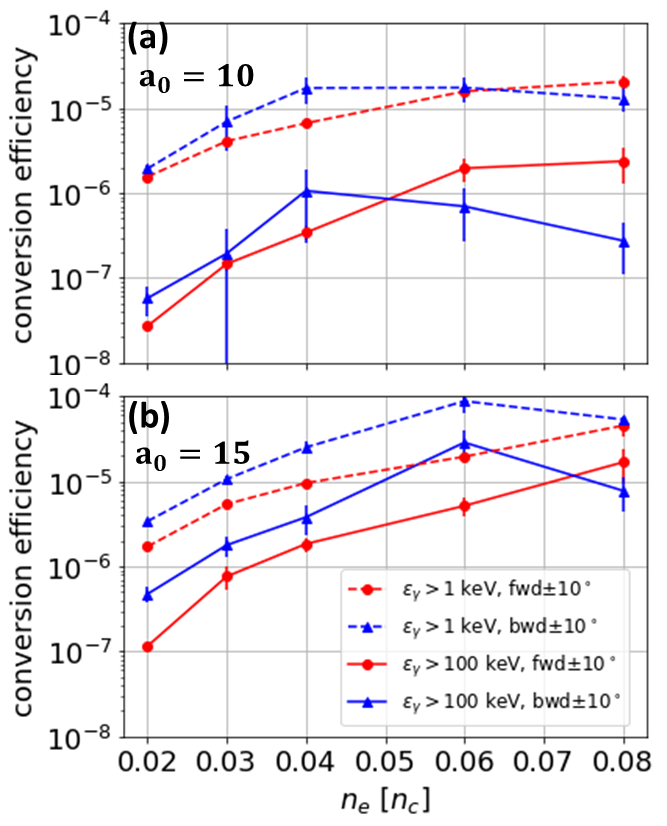}
    \caption{\label{fig:ce} Energy conversion efficiency for forward- and backward directed photons. (a)\&(b) plasma density scans for $a_0=10$ and $a_0=15$. Each marker is efficiency averaged over five 2D~PIC simulations. The error bars are the standard deviation for these five runs. 
} 
    \end{center}
\end{figure}

To conclude this section, we provide a parameter scan for the conversion efficiency of the laser energy into $\gamma$-rays. We define conversion efficiency as the ratio of the total energy in photons in a given range of $\varepsilon_{\gamma}$ to the total energy in the incident laser pulse. \Cref{fig:ce}(a) shows how the conversion efficiencies for $\varepsilon_{\gamma} > 1$~keV and for $\varepsilon_{\gamma} > 100$~keV scale with the plasma density. In our original simulation, the maximum electron density was $0.04 n_c$. The scan is performed by re-scaling the entire density profile using the same multiplier for each position along $x$, so the density profile is preserved. The horizontal axis in \cref{fig:ce}(a) is the peak density in each run. To account for the fact that the conversion efficiency might vary from run to run, we repeated each simulation (fixed physical parameters) five times using a different random seed in the PIC code. The markers show the average over these five runs and the error bars indicate the standard deviation. The main conclusion here is that the efficiency for the backward photons is comparable to the efficiency for the forward photons. Lowering the density below $0.04 n_c$ appears to be detrimental.

One last aspect that we want to highlight is the laser intensity dependence. All our simulations so far have used a laser with a peak field amplitude corresponding to $a_0 = 10$. We repeated the density scan for $a_0 = 15$ using the same procedure that we detailed in the previous paragraph. The result is shown in \cref{fig:ce}(b). As expected, the conversion efficiency goes up with the laser amplitude. The peak has shifted to higher density, which suggests that there is room for optimization. Moreover, the backward emission became more efficient at $a_0 = 15$. This observation together with the results presented in this section suggests that the backward emission warrants more attention from the community. 


\section{Summary and discussion} \label{sec: summary}

In this work, we have examined an interaction of a 700~fs long laser pulse with an underdense plasma of a finite length using 2D~PIC simulations. We found that, in addition to the forward emission of $\gamma$-rays, the plasma also emits in the backward direction. The emission is primarily concentrated at the density down-ramp. A slowly evolving longitudinal plasma electric field that builds up at the ramp is a key player for backward emission. It stops and re-accelerates in the backward direction moderate energy DLA electrons. The electrons emit as they interact with the laser pulse.   

We have assessed the backward emission using various metrics and found that it can be competitive with or even superior to the better known forward emission. The backward emission has a much smaller effective source size, which might be attractive to imaging applications. The energy conversion efficiency is comparable or even higher than that for the forward emission. The only potential downside is that the direction of the backward $\gamma$-ray beam fluctuates a lot more. Nevertheless, the these fluctuations are still less than the beam divergence angle.

Our analysis, presented in \cref{sec: pulse duration}, clearly indicates that using a longer laser pulse duration enhances backward $\gamma$-ray emission. However, there is likely an upper limit to the laser pulse duration's effectiveness for this enhancement. A possible limiting factor that requires examination is ion dynamics. The field generated by plasma electrons at the density down-ramp accelerates ions down the density gradient. This well-known effect tends to reduce field strength by decreasing space charge density. In our simulations, there is insufficient time for ion motion to manifest itself by affecting the field. Nevertheless, once a reduction in field strength becomes noticeable, it would diminish the ability to turn around and backward-accelerate electrons to high energies, thereby impacting the emission physics.

All our results are based on 2D rather than 3D simulations due to considerations of computational resources. The physics we presented is not specific to a 2D setup, so qualitative changes are unlikely when transitioning from 2D to 3D. However, it is well known that 2D PIC simulations tend to overestimate the strength of the sheath electric field—the electric field created by the energetic plasma electrons at the density down-ramp. One reason for this overestimation is that the electrons creating the field cannot expand into the third dimension. In the context of laser-driven ion acceleration, this overestimation of field strength leads to overestimates of ion energies and ion spectra. In our case, the overestimation of field strength likely results in higher energy backward electrons than one would expect in 3D. A direct implication is an overestimation of the backward emission. Adjusting the density gradient offers a potential path toward increasing the sheath field strength and boosting the backward $\gamma$-ray emission.

At the end of \cref{sec: source size}, we presented density scans for two different laser intensities/amplitudes. As we increased $a_0$, all other laser parameters were kept fixed, including the focal spot size. Recent publications suggest that laser focusing~\cite{tang.njp.2024} and the transverse size of the laser beam~\cite{babjak.prl.2024} can significantly impact electron acceleration via DLA. Therefore, it is conceivable that further optimization based on these findings could lead to a substantial increase in $\gamma$-ray emission.

\appendix

\section{PIC simulation parameters} \label{appe: PIC}

\Cref{table:PIC} provides detailed parameters of the 2D-3V PIC simulations presented in the manuscript. Simulations were carried out using the fully relativistic kinetic PIC code EPOCH~\cite{Epoch} (version 4.19.1). 

\begin{figure}[htb]
    \begin{center}
    \includegraphics[width=0.9\columnwidth,clip]{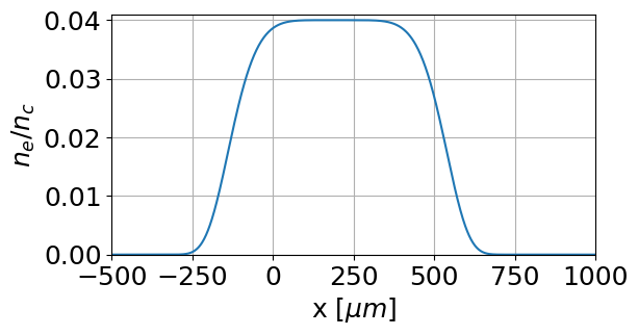}
    \caption{\label{fig:ne_profile} Initial profile of the normalized electron density in the longitudinal direction. 
} 
    \end{center}
\end{figure}

The plasma is initialized as a cold fully ionized helium plasma. The density profile along $x$ is shown in \cref{fig:ne_profile}. In the main simulation discussed in the paper, the peak density value is $0.04 n_c$. The laser is injected into the simulation box from the left ($x = -500~\micron$) and it propagates in the positive direction. We define $t=0$ as the time when the laser amplitude at $x = -500~\micron$ reaches its peak value. In other words, it is the time when the center of the laser pulse enters the simulation domain. 

Most of the analysis is based on the simulation where the normalized field amplitude is $a_0 = 10$ and the laser pulse duration is 700~fs. This simulation is compared with a simulation where the laser pulse duration is 200~fs. Additionally, we performed a parameter scan for a laser with higher intensity that corresponds to $a_0 = 15$.

\begin{table}[htb]
\centering
\begin{tabular}{ |p{4.2cm}|p{4cm}|  }
 \hline
 \multicolumn{2}{|c|} {Laser parameters} \\
 \hline \hline
 Normalized field amplitude & $a_0 = 10, 15$ \\
 \hline
 Peak intensity & $I_0 = 1.4, 3.1\times 10^{20}$ W/cm$^2$ \\
 \hline
 Wavelength & $\lambda_0 = 1.053$~$\micron$ \\
 \hline
 Focal plane of laser & $x=0$ $\micron$ \\
  \hline
 Transverse field profile in the focal plane & $\exp \left[ -\left( y/w_0  
 \right)^2 \right]$  \\
   \hline
 $w_0$ & $4.8~\micron$ \\
 \hline
 Temporal profile of laser electric field at left boundary & $\exp \left[ -\left( (t-t_0)/\tau 
 \right)^2 \right]$\\
 \hline
 $t_0$ & 980 fs \\
 \hline
 $\tau$ & 180, 640 fs\\
 \hline
  Focal spot size & \\
  \quad fwhm for field & $\Delta = 8~\micron$ \\
  \quad fwhm for intensity & 5.7 $\micron$ \\
 \hline
  Pulse duration & \\
  \quad fwhm for field  & $\tau_l =$ 700, 200 fs\\
  \quad fwhm for intensity & 495, 140 fs\\
  \hline
\end{tabular}
\bigskip

\begin{tabular}{ |p{4.2cm}|p{4cm}|  }
 \hline
 \multicolumn{2}{|c|}{Target parameters} \\
 \hline \hline
  Electron density profile along $y$ & uniform for $|y| \leq 45~\micron$ \\
 \hline
 Electron density profile along $x$ & $n_0  \exp \left[ - \left(\frac{x-200~\micron}{350~\micron}\right)^6 \right] $\\
 \hline
 Maximum electron density & $n_0 = 0.02 - 0.08n_\text{c}$ \\ 
\hline
 Composition & $\rm{He}^{2+}$ and $\rm{e}^{-}$ \\
 \hline
\end{tabular}

\bigskip

\begin{tabular}{ |p{4.2cm}|p{4cm}|  }
 \hline
 \multicolumn{2}{|c|}{{Other parameters}} \\
 \hline \hline
 Simulation box & $x=[-500,1000]$ $\micron$;\\ 
 & $y=[-50,50]$ $\micron$\\
 \hline
 Spatial resolution & 30 cells per $\micron$ in $x$\\
 & 30 cells per $\micron$  in $y$\\
 \hline
 Macro-particles per cell & 4 for $\rm{e}^{-}$ \\
 & 2 for $\rm{He}^{2+}$ \\
 \hline
\end{tabular}

\caption{2D PIC simulation parameters.}
  \label{table:PIC}
  
\end{table}

\section{Results from multiple runs} \label{appe: multiple_run}

\begin{figure}[htb]
    \begin{center}
    \includegraphics[width=1\columnwidth,clip]{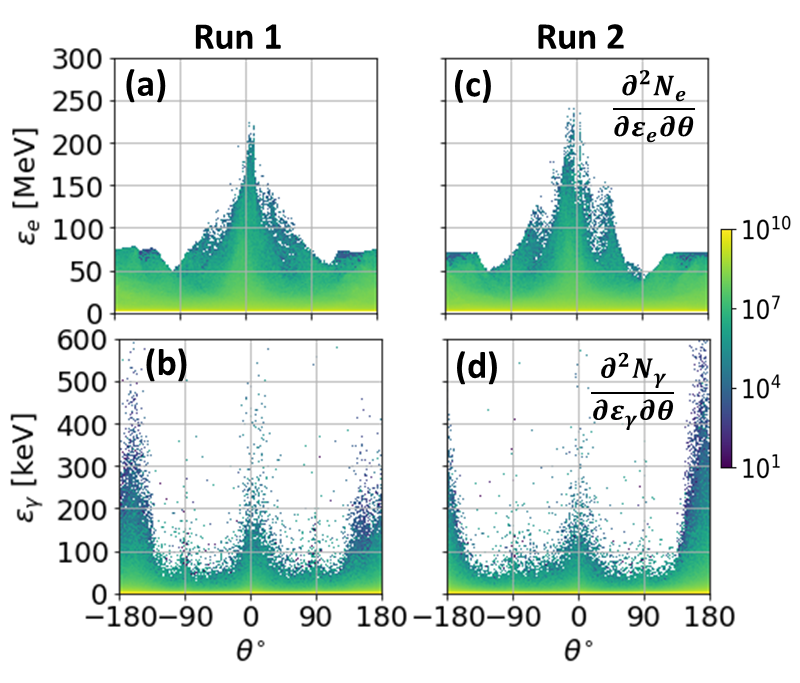}
    \caption{\label{fig:multiple_runs} Two additional simulations performed using the same parameters as those used to generate \cref{fig:field_spectrum}. These simulations use random seeds that are different from that used the original simulation. 
    Shown are angle-resolved electron spectra, (a)\&(c), and photon spectra, (b)\&(d), during the propagation of the 700~fs laser pulse along the density down-ramp ($t = 4$~ps). For electrons, the color is the number of electrons per MeV per degree, i.e.  $1/(\rm{MeV}\,\rm{deg}^{\circ})$. For photons, the color is the number of photons per keV per degree, i.e. $1/(\rm{keV}\,\rm{deg}^{\circ})$.} 
    \end{center}
\end{figure}  

The key features of our simulations are reproducible, but there are fluctuations that are caused by plasma instabilities. To explore the impact of these fluctuations, we performed two additional runs. These runs use the same physical parameters that were used to generate the spectra in \cref{fig:field_spectrum}. The only difference is that we changed the random seed used by the PIC code. The resulting spectra for the two new runs are shown in \cref{fig:multiple_runs}. One can see that the two new runs also have backward-accelerated electrons and they feature strong backward $\gamma$-ray emission.

The code uses random numbers to place the macro-particles during the initialization and to generate photons. In our case, electron recoil during the emission is very weak because of the small $\chi$. It is then unlikely that the emission itself can produce noticeable fluctuations. Most likely, the fluctuations that we see in the spectra result from differences in initial particle placement.


\section*{Acknowledgments}

This material is based upon work supported by the Department of Energy National Nuclear Security Administration under Award Number DE-NA0004030. K.T. was supported by the National Science Foundation–Czech Science Foundation partnership (NSF Grant No. PHY-2206777). 
Simulations were performed with EPOCH (developed under UK EPSRC Grants No. EP/G054940/1, No. EP/G055165/1, and No. EP/G056803/1) using HPC resources provided by TACC at the University of Texas. 

\section*{References}
\bibliography{Collection}


\end{document}